\documentclass[twocolumn,preprintnumbers,amsmath,amssymb]{revtex4}

\usepackage{graphicx}
\usepackage{dcolumn}
\usepackage{bm}
\usepackage{times}
\usepackage{amsmath}
\usepackage{amsfonts}
\usepackage{amssymb}

\begin{document}

\title{Optical signatures of quantum dot excitons in carbon nanotubes}

\author{Matthias S. Hofmann$^{*}$ \footnote[0]{\scriptsize$^{\rm *}$ These authors have contributed equally to this work.}}
\author{Jan T. Gl\"uckert$^{*}$}
\author{Alexander H\"ogele}

\affiliation{Fakult\"at f\"ur Physik and Center for NanoScience
(CeNS), Ludwig-Maximilians-Universit\"at M\"unchen,
Geschwister-Scholl-Platz 1, D-80539 M\"unchen, Germany}

\date{\today}

\begin{abstract}
We report optical studies of quantum dot excitons in individual
suspended carbon nanotubes at cryogenic temperatures. Narrow
optical linewidths, strongly suppressed spectral wandering, and
photoluminescence lifetimes in the range of nanoseconds emerge as
key signatures of exciton localization. We infer exciton quantum
dot formation with a characteristic length of a few exciton Bohr
radii. Localization inhibits exciton diffusion and protects the
exciton from dephasing by structural or environmental
inhomogeneities as well as from exploring nonradiative quenching
sites. In consequence, quantum dot excitons in carbon nanotubes
exhibit intrinsic radiative lifetimes, long coherence times and a
quantum yield of $100\%$. Our results underpin the potential of
carbon nanotube excitons for both fundamental studies and
applications that scale advantageously with enhanced spectral
resolution and coherence.
\end{abstract}

\maketitle

Semiconducting single-walled carbon nanotubes (CNTs) exhibit
absorption \cite{OConnell2002} and emission \cite{Bachilo2002} in
the near-infrared as a consequence of excitonic
\cite{Wang2005,Maultzsch2005} creation and recombination pathways.
Residing entirely on the nanotube surface and mobile along the
axis on length scales that at room temperature far exceed the
exciton Bohr radius \cite{Lefebvre2006,Cognet2007,Luer2009}
nanotube excitons are highly sensitive to variations in the CNT
structure as well as the immediate surrounding. This, on the one
hand, is very attractive for CNT-based applications in optical
sensing \cite{Barone2005,Withey2012}. On the other hand it renders
CNT excitons susceptible to environmental dephasing
\cite{Duque2009} or quenching at nonradiative defect sites
\cite{Cognet2007}. In consequence, extrinsic effects currently
limit in part our quantitative understanding of the photophysical
properties of CNTs which is key to the development of novel
nanotube-based optoelectronic devices.

Here we demonstrate that exciton localization in as-grown
suspended nanotubes facilitates the study of intrinsic optical
properties of CNTs. As a consequence of quantum dot (QD) exciton
formation we found photoluminescence (PL) lifetimes in the range
of nanoseconds at the temperature of liquid helium ($4.2$~K) as
predicted by theory for the intrinsic radiative lifetime of CNT
excitons \cite{Perebeinos2004,Perebeinos2005,Spataru2005}.
Moreover, the PL spectra are as narrow as $40~\mu$eV and free of
spectral wandering, identifying $10$~ps time scale as a lower
bound to the exciton coherence time. Alike monolayer-fluctuation
QDs in narrow semiconductor quantum wells \cite{Hess1994} or QDs
formed by self-assembly \cite{Leonard1993}, QDs in our CNT
material occur unintentionally. In addition to spectrally narrow
emission lines QD excitons in CNTs share other remarkable
properties with their celebrated zero-dimensional counterparts in
compound semiconductors such as characteristic resonances in
photoluminescence excitation (PLE), spectral signatures of
exciton-phonon interaction and strong photon antibunching
\cite{Hoegele2008}.

\begin{figure}[t]
\includegraphics[scale=0.9]{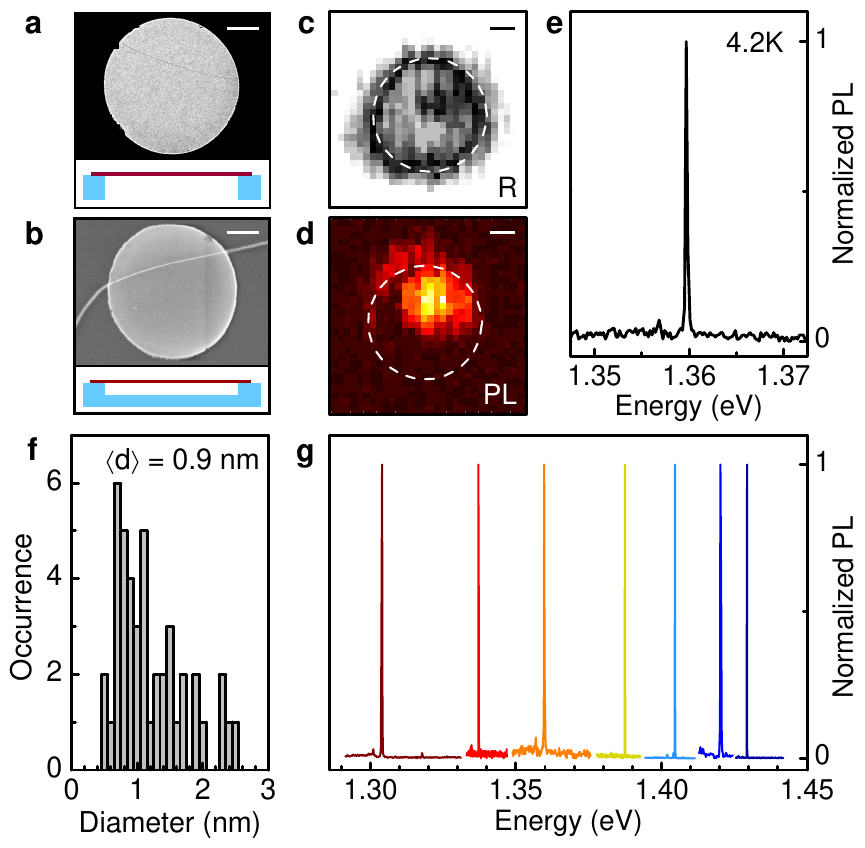}
\caption{({\bf a}) Transmission and ({\bf b}) scanning electron
micrographs and cross-section schematics of sample regions with
carbon nanotubes suspended over a hole and a SiO$_2$ crater of
$200$~nm depth, respectively. False-color maps recorded in
reflection ({\bf c}) and PL ({\bf d}) for a single carbon nanotube
with a sharp emission spectrum ({\bf e}). Dashed white circles in
({\bf c}) and ({\bf d}) indicate the perimeter of the circular
crater. Scale bars in ({\bf a})\,-\,({\bf d}) are $500~$nm. ({\bf
f}) The diameter distribution yields an average diameter of
$\langle \mathrm{d} \rangle=0.9$~nm. ({\bf g}) Exemplary PL
spectra in the energy range of $1.300-1.450$~eV are represented in
different colors, all exhibiting narrow linewidths of
$40-100~\mu$eV at $4.2$~K.} \label{fig1}
\end{figure}

Our surprising findings are a result of the interplay between
rational sample design and accidental exciton localization. A
central aspect of our sample design, inspired by results from
electron transport spectroscopy in clean CNTs \cite{Cao2005}, was
the use of as-grown suspended CNTs without any post-processing.
Transmission and scanning electron micrographs in Fig.~\ref{fig1}a
and \ref{fig1}b show representative images of suspended CNTs on
our sample. The carrier substrate, Si$_3$N$_4$ coated with
SiO$_2$, has two distinct regions with holes and craters of
$2~\mu$m diameter (schematics in Fig.~\ref{fig1}a and b). Our
chemical vapor deposition (CVD) method, unspecific to spatial
catalyst positioning, yielded a homogeneous CNT density over the
entire sample of mm dimensions. In this study we present
measurements on CNTs suspended over craters as in
Fig.~\ref{fig1}b \cite{SM}.

A representative confocal PL map of such a crater is shown in
Fig.~\ref{fig1}d: intense and faint PL signals were detected in
the inner part and at the outer perimeter of the crater,
respectively. The spatial extent of the crater was mapped out with
raster-scan reflection imaging (the reduced laser reflection by
the crater ground in Fig.~\ref{fig1}c is due to its
out-of-focal-plane displacement). The PL intensity displayed a
pronounced antenna effect \cite{SM}, a characteristic feature of
individual CNTs \cite{Hartschuh2003}. Spectral dispersion of the
luminescence revealed a remarkably narrow emission line
(Fig.~\ref{fig1}e) centered at $1.360$~eV and confined to $1-3$
CCD-pixels of $40~\mu$eV spectral width. This feature of a
resolution-limited linewidth was consistently observed for all
CNTs studied in the emission window $1.300-1.450$~eV with
representative PL spectra displayed in Fig.~\ref{fig1}g. The
spread in emission energies is related on the one hand to
different chiralities present in our CVD-material with a mean
diameter of $\langle \mathrm{d} \rangle=0.9$~nm
(Fig.~\ref{fig1}f), and to dissimilarities between CNTs of the
same chirality on the other hand. CNTs of five different
chiralities, $(5,4), (6,2), (6,4), (8,3)$ and $(9,1)$, are likely
to emit at $4.2$~K into the spectral window of Fig.~\ref{fig1}g
\cite{SM}.

\begin{figure}[t]
\includegraphics[scale=0.9]{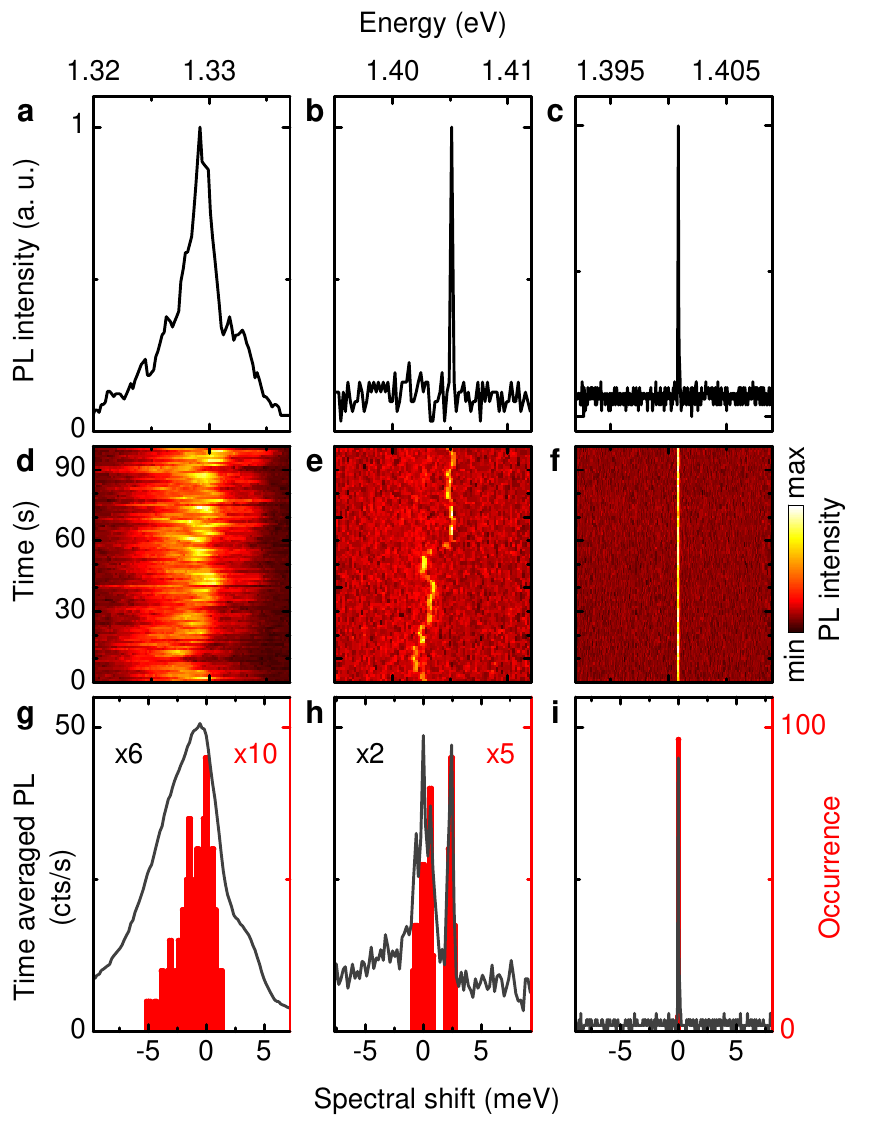}
\caption{Photoluminescence spectra measured in $1$~s for ({\bf a})
a micelle-encapsulated CoMoCat-nanotube on SiO$_2$, ({\bf b}) a
single CVD-nanotube on SiO$_2$, and ({\bf c}) a single
CVD-nanotube suspended over a SiO$_2$ crater. ({\bf d})\,-\,({\bf
f}) Corresponding time-traces of successive PL spectra with $1$~s
integration time in false-color representation. ({\bf
g})\,-\,({\bf i}) Histograms of the PL maximum position (red bars)
and time averaged PL intensity (black spectra) derived from the
respective time-traces are shown as a function of the spectral
shift. All spectra were measured at $4.2$~K.} \label{fig2}
\end{figure}

Narrow optical linewidths at cryogenic temperatures are untypical
for CNTs \cite{Lefebvre2004,Htoon2004,Kiowski2007b,Hoegele2008}.
For reference we studied nanotubes obtained by the same growth
procedure in contact with SiO$_2$ as well as commercial
CoMoCat-nanotubes with comparable diameters \cite{Bachilo2003}
encapsulated in SDS and dispersed on SiO$_2$ \cite{SM}. Under
similar experimental conditions CoMoCat-nanotubes typically
exhibited asymmetric PL profiles
\cite{Htoon2004,Hoegele2008,Galland2008} with linewidths of $\sim
1~$meV (Fig.~\ref{fig2}a), one order of magnitude broader than our
CVD-grown nanotubes in $1$~s integration time (Fig.~\ref{fig2}b
and c). While on longer timescales of the experiment the linewidth
of the time-averaged PL broadened even further for reference CNTs
(time averaged PL spectra in Fig.~\ref{fig2}g and h), suspended
CNTs showed no change in the emission profile irrespective of the
integration time (effective integration time of $100$~s in
Fig.~\ref{fig2}i and up to several days of observation). PL time
traces in Fig.~\ref{fig2}d - f reveal that spectral wandering is
responsible for the broadening of the optical linewidth on meV
energy scale, in accord with previous reports
\cite{Htoon2004,Kiowski2007b,Hoegele2008,Matsuda2008}. At the same
time it also accounts for apparent fine-structure
\cite{Lefebvre2004,Matsuda2008} and to some extent for the
asymmetry in the spectral profile (histograms in Fig.~\ref{fig2}g
- I): an initially narrow spectrum of a CVD-nanotube in contact
with SiO$_2$ (Fig.~\ref{fig2}b) developed in the course of
integration (corresponding to time averaging of the PL trace in
Fig.~\ref{fig2}e) an asymmetric peak accompanied by a satellite
(Fig.~\ref{fig2}h). For suspended CNTs, these features of spectral
wandering are entirely absent within the spectral resolution and
on all time scales of our experiment (Fig.~\ref{fig2}c, f, i).

\begin{figure}[t]
\includegraphics[scale=0.9]{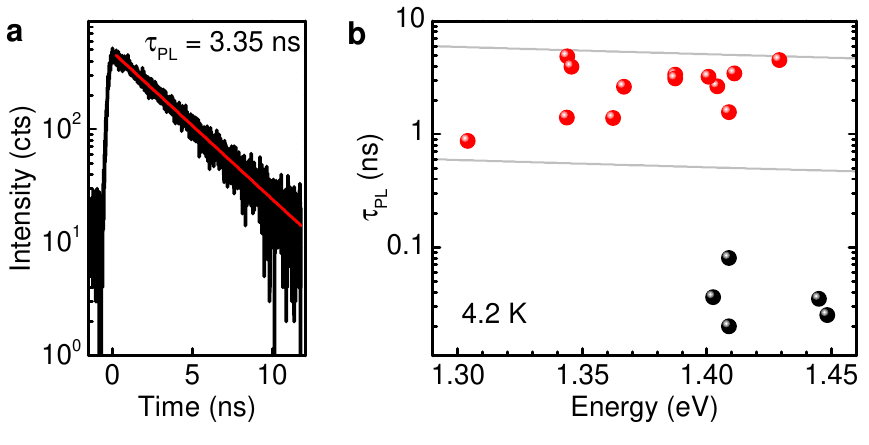}
\caption{({\bf a}) Monoexponential PL decay of a single suspended
nanotube at $4.2$~K with a decay time of
$\tau_{\mathrm{PL}}=3.35$~ns. ({\bf b}) The PL lifetimes of
suspended nanotubes are in the range of $1-5$~ns (red circles),
one order of magnitude longer than for surfactant-encapsulated
CoMoCat-nanotubes on SiO$_2$ (black circles). Limits to the
radiative lifetime of excitons localized within their Bohr radius
and free excitons are represented by the upper and lower grey
lines, respectively.}\label{fig3}
\end{figure}

Another remarkable signature of suspended CNTs was revealed by
time-resolved PL spectroscopy. For all suspended tubes we observed
monoexponential PL decay with timescales above one nanosecond. A
PL decay trace for a single CNT with a decay time of $3.35$~ns
is shown in Fig.~\ref{fig3}a. Nanoseconds decay times, one order
of magnitude longer than in the CoMoCat reference material (red vs
black circles in Fig.~\ref{fig3}b), are in stark contrast to all
previous reports. Room-temperature experiments determined both
monoexponential \cite{Gokus2008} and biexponential
\cite{Berciaud2008,Miyauchi2009,Gokus2010} decays depending on the
material quality with decay times of the order of tens of ps.
Similar results were obtained at low temperatures
\cite{Hagen2005}. Consensus between $10-100$~ps PL lifetimes
observed experimentally and intrinsic exciton lifetimes predicted
by theory \cite{Perebeinos2004,Perebeinos2005,Spataru2005} was
achieved by assuming rapid nonradiative decay channels arising
from defects.

\begin{figure}[t]
\includegraphics[scale=0.9]{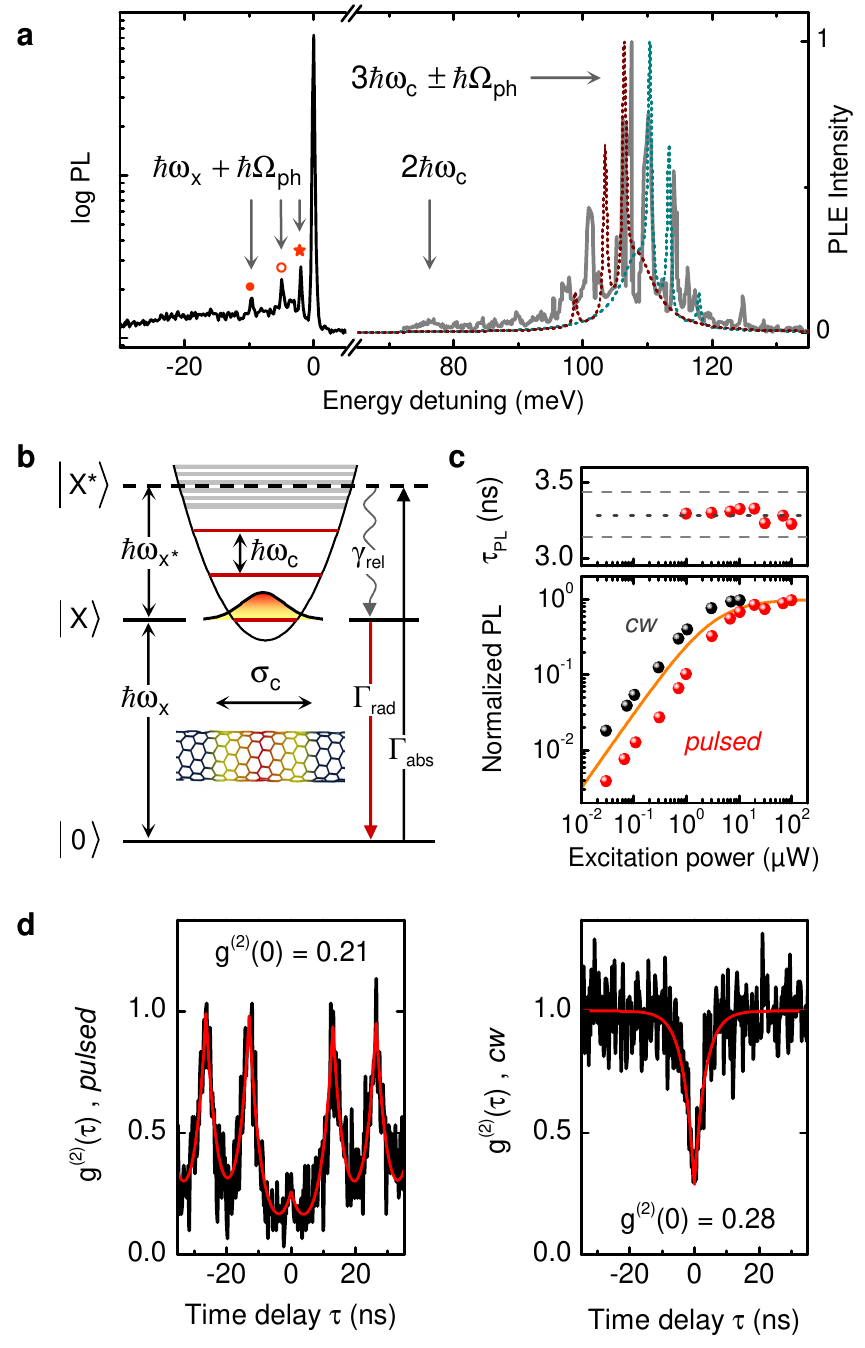}
\caption{({\bf a}) Both the PL (left) and the PLE (right) spectra
are modulated by phonons with energies $\hbar
\Omega_{\mathrm{ph}}$. Resonances in the excitation arise at $2$
and $3\hbar\omega_{\mathrm{c}}$. Dashed lines are model Raman
spectra of the zero-phonon absorption line at
$3\hbar\omega_{\mathrm{c}}$ with $6~$meV width and phonon replicas
labelled by $\star$, $\circ$ and $\bullet$. ({\bf b}) Schematics
of an optical cycle of a harmonically confined exciton with
relevant states, energies and rates. ({\bf c}) PL lifetime (upper
panel) and saturation (lower panel) of the CNT in Fig.~\ref{fig3}a
as a function of laser power (red and black data correspond to cw
and pulsed excitation). The solid line shows the saturation of a
three level system with a quantum yield of $100\%$. ({\bf d})
Second-order photon correlation function for the same nanotube
under pulsed (left) and cw (right) excitation with pronounced
antibunching at zero time-delay (red lines were calculated using
$\tau_{\mathrm{PL}}=3.35$~ns).} \label{fig4}
\end{figure}

It is unlikely that our CNTs are entirely free of defects. Even if
they were perfect in crystalline structure and free of surface
adsorbates, the exciton should diffusively explore the nanotube
and eventually encounter the substrate supports at the ends of the
suspended segment of $2~\mu$m length at maximum. We argue
therefore that the excitons in our CNTs are localized in QDs.
Being immobile they do not encounter nonradiative quenching sites
and therefore exhibit intrinsic exciton lifetimes. At the same
time localization ensures narrow emission profiles by protecting
the exciton from diffusively exploring inhomogeneities along the
tube axis which would give rise to line broadening via dephasing.
Taking the resolution limit as a conservative value for the total
linewidth we arrive at dephasing times above $10$~ps. Unlike
diffusive excitons in CNTs, localized QD excitons exhibit both
long decay and dephasing times as key signatures of their optical
characteristics.

Taking the PL decay time as the intrinsic exciton lifetime allows
us to estimate the strength of the confinement potential
\cite{SM}. Localization limits the exciton coherence length to the
confinement length $\sigma_{\mathrm{c}}$ (Fig.~\ref{fig4}b) that
we estimate to $1-5$ exciton Bohr radii $\sigma_{\mathrm{X}}\sim
1.2$~nm \cite{Capaz2006} on the basis of PL lifetimes in the range
of $1-5~$ns and theoretically calculated oscillator strengths
\cite{Perebeinos2004,Perebeinos2005,Spataru2005}. For the QD
exciton in Fig.~\ref{fig3}a, $\tau_{\mathrm{PL}}=3.35~$ns implies
a harmonic level spacing of $38.5~$meV. This number was confirmed
with PLE spectroscopy (grey spectrum in Fig.~\ref{fig4}a): the PL
emission was enhanced for laser energy detunings of
$\hbar\omega_{\mathrm{X}^*}=75.7$~meV and $108.4$~meV above the
lowest exciton state $|\mathrm{X}\rangle$, consistent with
resonances at $2$ and $3\hbar \omega_{\mathrm{c}}$ and weak
anharmonicity. Both the PL (left graph in Fig.~\ref{fig4}a) and
the PLE spectra exhibited modulation by phonons with energies
$\hbar \Omega_{\mathrm{ph}}$ in the meV range. Clearly, different
phonons are responsible for the modulation of the ground state
$|\mathrm{X}\rangle$ and the third excited state
$|\mathrm{X}^*\rangle$ that is of opposite symmetry: our simple
model of the absorption resonance (Stokes and Anti-Stokes Raman
spectra in Fig.~\ref{fig4}a calculated as a convolution of the
zero-phonon absorption line at $3\hbar \omega_{\mathrm{c}}$ with
$6.1~$meV linewdith and $0.5~$meV broad phonon replica with
energies $2.0~$meV, $5.0~$meV and $9.5~$meV as found in PL
emission) fails to reproduce the PLE spectrum. The rich structure
related to exciton-phonon interaction requires a more detailed
analysis.

Remarkably, our quasi-resonant excitation scheme yields the
saturation response of a three-level system (Fig.~\ref{fig4}c). In
time-resolved PL we found no evidence for Auger decay typically
responsible for saturation effects in the CNT emission
\cite{Murakami2009}. Despite saturation above $5~\mu$W excitation
powers the PL lifetime remained monoexponential within the
temporal resolution of our detector (Fig.~\ref{fig4}c) and without
signatures of an emerging secondary decay. At the same time photon
correlation results with pronounced antibunching
(Fig.~\ref{fig4}d) exclude multiexciton recombination. Instead,
the saturation response is quantitatively captured with a model
three-level system \cite{SM}. The saturation function shown as
solid line in Fig.~\ref{fig4}c was obtained without free
parameters, implying for QD excitons in CNTs a quantum yield of
$100\%$ and no indication for shelving in the lowest-lying dark
exciton states \cite{Perebeinos2005,Spataru2005}.

Our results establish QD excitons as a new regime of CNT optics.
In our samples exciton localization was obtained by chance. There
is no fundamental reason, however, that precludes QD formation by
design. Electrostatic traps are commonly used in CNT transport
experiments to define and vary localization boundaries of
electrons or holes, a strategy also applicable to neutral excitons
\cite{Wilson-Rae2009}. Chemical functionalization of nanotube ends
or structural modification on the atomic scale as used to assemble
designer graphene \cite{Gomes2012} are alternative strategies to
control the position and the extent of exciton localization.
Paired with the exceptional mechanical properties of CNTs our
results also stimulate novel experiments in nano-optomechanics.

The research was funded by the German Excellence Initiative via
the Nanosystems Initiative Munich (NIM), with financial support
from the Center for NanoScience (CeNS) and LMUexcellent. We thank
J.~P.~Kotthaus for continuous support, S.~Stapfner, C.~Bourjau,
R.~Dehmel, and F.~Storek for assistance with nanotube synthesis
and sample fabrication. We acknowledge C.~Sch\"onenberger and
M.~Weiss for introducing to us in the early stage of the project
their CNT synthesis at the University of Basel, Switzerland. A.H.
thanks A.~Imamoglu, P.~Maletinsky, V.~Perebeinos, S.~Rotkin,
A.~Srivastava, and I.~Wilson-Rae for useful discussions.




\cleardoublepage

\setcounter{figure}{0} \setcounter{equation}{0}

\hspace{35pt} {\Large{\textbf{Supplementary material}}}


\section{Experimental details}

\subsection{Carbon nanotube samples}

Chemical vapor deposition (CVD) was used to synthesize carbon
nanotubes (CNTs) in a standard CVD furnace. The growth was
assisted by a bimetallic FeRu catalyst. Catalyst particles were
deposited on carrier substrates from the FeRu suspension either by
spin or by drop coating. The samples were heated in the CVD
furnace to $800-850^\circ$C in an Ar/H$_2$ ($95\%/5\%$) gas
mixture. The growth procedure was optimized to yield a CNT density
below $1~\mu$m$^{-2}$ and narrow-diameter nanotubes with a mean
diameter below $1$~nm. Diameter distributions were obtained by
tapping-mode atomic force microscopy on samples with CNTs on
SiO$_2$. Samples with as-grown CNTs in contact with SiO$_2$ were
used as reference material for density and diameter distributions
as well as in cryogenic spectroscopy.

\begin{figure*}[!h]
\hspace{-10pt}
\includegraphics[scale=1]{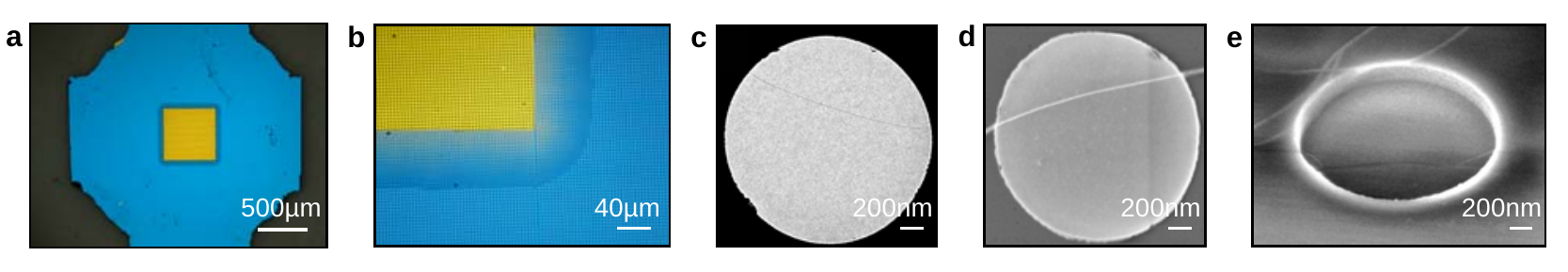}
\caption{({\bf a}), ({\bf b}) Optical microscopy images of a TEM
grid used as a carrier substrate. A Si$_3$N$_4$ membrane with
regular hole patterns lies on top of a silicon frame (blue region
of the TEM grid). In the central area (yellow region of the TEM
grid window) the holes provide for the suspension of nanotubes
free of substrate underneath. ({\bf c}) TEM image of a carbon
nanotube suspended over a hole. ({\bf d}), ({\bf e}) SEM images of
carbon nanotubes suspended over a SiO$_2$ crater and in partial
contact with the ground of the crater, respectively.}
\label{sample}
\end{figure*}

Suspended CNTs were synthesized on commercial grids typically used
in transmission electron microscopy (TEM). The specific grid used
in the experiments was perforated with holes of $2~\mu$m diameter
and a pitch of $4~\mu$m (Fig.~\ref{sample}a) and coated with
$100~$nm of SiO$_2$ by plasma enhanced CVD. Fig.~\ref{sample}a and
\ref{sample}b show optical micrographs of the sample for two
different magnifications. The grids provided for suspension of
nanotubes in the region where the hole-perforated Si$_3$N$_4$
membrane was not supported by the underlying SiO$_2$ frame; this
region appears yellowish in Fig.~\ref{sample}a, b. In this region
fully suspended nanotubes were identified by TEM imaging
(Fig.~\ref{sample}c). In regions where the perforated membrane was
in contact with the underlying carrier frame (blue regions in
Fig.~\ref{sample}a and \ref{sample}b) we identified with scanning
electron microscopy (SEM) CNTs suspended over the full diameter of
the crater (Fig.~\ref{sample}d) as well as CNTs that were
suspended only near the walls of the crater (Fig.~\ref{sample}e).

Commercial CoMoCAT-nanotubes (SouthWest NanoTechnologies)
encapsulated in sodium dodecylbenzenesulfonate (SDS) were
dispersed out of an aqueous suspension on SiO$_2$ substrates.

\subsection{Cryogenic spectroscopy}

For cryogenic spectroscopy we used a home-built confocal
microscope with a full-width-half-maximum ($\mathrm{FWHM}$)
diffraction-limited optical spot size of $\mathrm{FWHM}\simeq
\lambda$. The two-lens imaging system was not compensated for
chromatic aberrations. The microscope was operated in a bath
cryostat at temperatures of liquid nitrogen ($77$~K) or liquid
helium ($4.2$~K).

\begin{figure*}[!b]
\hspace{-35pt}
\includegraphics[scale=1]{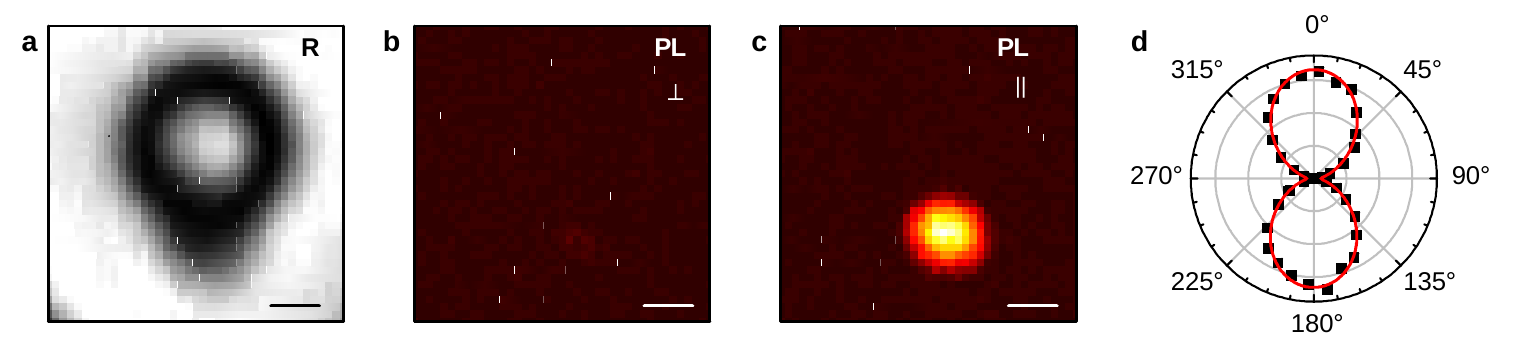}
\caption{({\bf a}) False-color map of a crater in reflection.
({\bf b}) PL map of the same crater with laser polarization
perpendicular and ({\bf c}) parallel to the nanotube axis (scale
bars are $1~\mu$m). ({\bf d}) Polar plot of the PL intensity as a
function of the angle between the laser polarization axis and the
the nanotube axis. The red solid line is a fit with a
$\cos^2$-dependence.} \label{polarization}
\end{figure*}

The sample was characterized by scanning a finite area of a few
$\mu\mathrm{m}^2$ and recording either the reflected signal of a
laser diode at $905$~nm wavelength (Fig.~\ref{polarization}a) or
the PL intensity (Fig.~\ref{polarization}b and
\ref{polarization}c). For PL excitation a Ti:Sapphire laser in
continuous wave (cw) mode was tuned between $730$~nm and $850$~nm.
Strong antenna effect in the PL intensity
(Fig.~\ref{polarization}b, c, d) was characteristic for sample
regions with individual CNTs. The PL was dispersed in a standard
monochromator ($500$~mm focal length) and detected with a nitrogen
cooled silicon-CCD (with $40~\mu$eV spectral width of one CCD
pixel).

For time-resolved PL and pulsed photon correlation in a standard
Hanbury-Brown and Twiss setup the Ti:sapphire laser was operated
in the fs-pulsed mode ($130$~fs pulse width, $76.0$~MHz repetition
rate). PL was recorded by avalanche photodiodes (APDs) with a
temporal resolution of $300$~ps. A streak-camera with $7$~ps
temporal resolution was used for PL lifetime measurements of
micelle-encapsulated CoMoCAT nanotubes.

\begin{figure*}[!b]
\hspace{-40pt}
\includegraphics[scale=1.1]{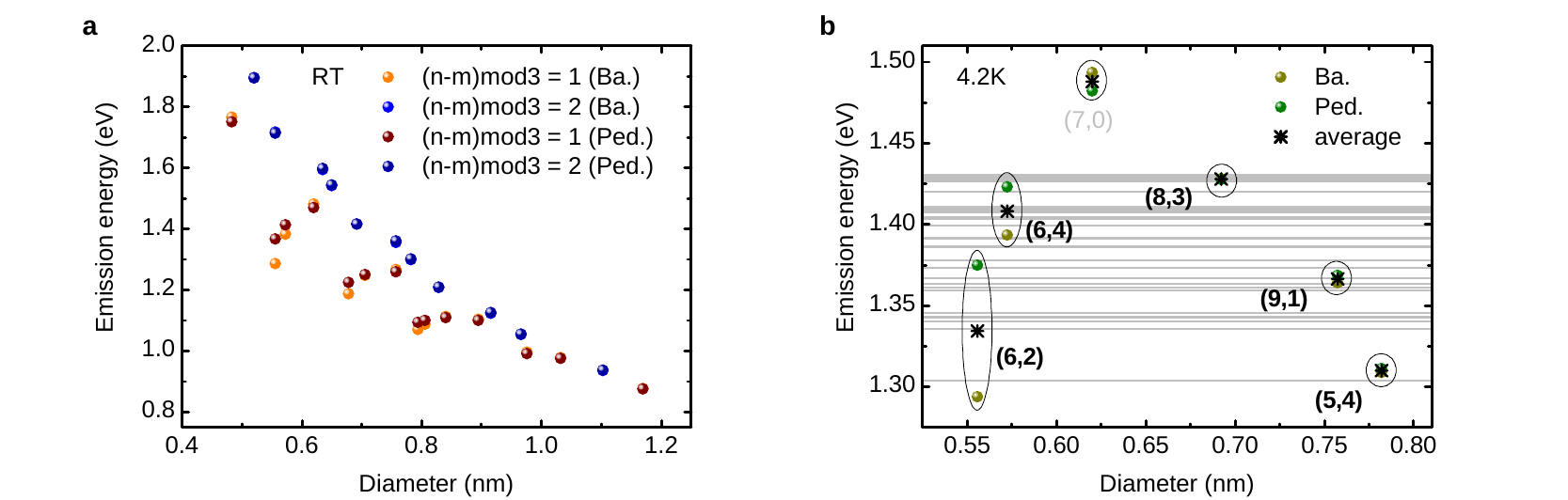}
\caption{({\bf a}) Calculated emission energies E$_{11}$ of CNTs
with different chiralities at room temperature (RT) as a function
of diameter according to Bachilo {\it et al.} (Ba.)
\cite{Bachilo2002} and Pedersen (Ped.) \cite{Pedersen2004}. ({\bf
b}) Emission energies of small-diameter nanotubes at $4.2$~K
taking into account temperature dependent energy shift according
to Capaz {\it et al.} \cite{Capaz2005}. Gray lines represent the
experimentally observed emission energies. Stars indicate emission
bands presumably related to the emission from CNTs with five
different chiralities: $(5,4), (6,2), (6,4), (8,3)$, and $(9,1)$.}
\label{E-vs-d}
\end{figure*}

Emission energies of all suspended CNTs studied in this work at
$4.2$~K are represented in Fig.~\ref{E-vs-d}b by gray horizontal
lines. In order to correlate the nanotube emission with chirality,
we calculated in a first step room temperature emission energies
according to Bachilo {\it et al.} \cite{Bachilo2002} and Pedersen
\cite{Pedersen2004}. The resulting emission energies for the
$E_{11}$-transitions are depicted in Fig.~\ref{E-vs-d}a. In a
second step the emission energies were corrected for temperature
related spectral shifts according to Capaz {\it et al.}
\cite{Capaz2005}. The results for the relevant spectral range
$1.30-1.44$~eV are summarized in Fig.~\ref{E-vs-d}b. From
comparison of calculated emission energies with experimental data
we limit the relevant chiralities to $(5,4), (6,2), (6,4), (8,3)$,
and $(9,1)$. Further corrections to the emission energy due to
strain or details of the confinement potential were not
considered.

\section{Theoretical modelling}

\subsection{Exciton lifetimes}

Following Perebeinos {\it et al.} \cite{Perebeinos2005} we
calculate the intrinsic radiative lifetime of CNT excitons as the
inverse of the radiative decay rate


\begin{equation}
\frac{1}{\tau} = \Gamma_{\mathrm{rad}}= \frac{n_r e^2 E_{11}^2
f}{2\pi \epsilon_0 m_0 \hbar^2 c^3}, \label{eqn-tau}
\end{equation}

\noindent with the dielectric permittivity $\epsilon_0$, the free
electron mass $m_0$, the speed of light $c$, and the refractive
index $n_{r}=\sqrt{\epsilon}$, with $\epsilon=1.846$ for a CNT in
vacuum \cite{Capaz2006}. For a CNT with chirality $(n,m)$ and
diameter $d=(a_{cc}/\pi) \sqrt{3(n^2+m^2+nm)}$
($a_{cc}=0.144~\mathrm{nm}$ is the the C--C bond length)
\cite{SaitoBook} the exciton emission energy is approximated by
$E_{11}=0.84~\mathrm{eV nm}/d$ and the oscillator strength per
carbon atom by $f=0.014~\mathrm{eV}^{-1} E_{11} \pi \, d \,
\sigma_{\mathrm{c}}/A$  \cite{Perebeinos2004}. Here
$\sigma_{\mathrm{c}}$ is the coherence length given by the
confinement length, and $A=3\sqrt{3}/4\, a_{cc}^2$ is the area per
carbon atom. The exciton Bohr radius $\sigma_{\mathrm{X}}$ is
calculated according to Ref.~\cite{Capaz2006}. The values for the
intrinsic exciton lifetimes obtained from Eq.~\ref{eqn-tau} are in
good agreement with \textit{ab initio} calculations by Spataru
{\it et al.} \cite{Spataru2005}, see Table~\ref{table_tau}.

\begin{table}[h]
\caption{Comparison of intrinsic radiative lifetimes of free
excitons $\tau$(0) obtained with \textit{ab initio} calculations
by Spataru {\it et al.} \cite{Spataru2005} and from
Eq.~\ref{eqn-tau} following Perebeinos {\it et al.}
\cite{Perebeinos2004,Perebeinos2005}.}
    \centering
    \vspace{20pt}
        \begin{tabular}{    p{2.5cm}    p{2.5cm}    p{2.5cm}    }
\hline\hline
                 &  Ref. \cite{Spataru2005}:    & Ref. \cite{Perebeinos2005}: \\
$(n,m)$    &    $\tau$(0) (ps)                  & $\tau$(0) (ps) \\
\hline
(7,0)            & 12.8                                         & 8.6  \\
(8,0)            & 8.1                                          & 9.8  \\
(10,0)     & 19.1                                           & 12.2 \\
(11,0)       & 14.3                                         & 13.5  \\
\hline\hline
        \end{tabular}
        \label{table_tau}
\end{table}

\noindent The energy level spacing $\hbar\omega_{\mathrm{c}}$ of a
one-dimensional harmonic confinement potential with confinement
length $\sigma_{\mathrm{c}}$ is obtained from

\begin{equation}
\hbar\omega_{\mathrm{c}}=\frac{\hbar^2}{M_{\mathrm{X}}\sigma_{\mathrm{c}}^2}
\label{eqn-spacing}
\end{equation}

\noindent using the exciton mass $M_{\mathrm{X}}=m_e^*+m_h^*$, and
$m_e^*$, $m_h^*$ calculated according to Ref.~\cite{Jorio2005}.

\subsection{Saturation of a three-level model system}

With the definitions given in the main text we obtain the
saturation behavior of a three-level system (set up by the crystal
ground state $|0 \rangle$, and the exciton ground and excited
states, $| \mathrm{X}\rangle$ and $| \mathrm{X}^*\rangle$,
respectively) by solving the coupled rate equations

\begin{eqnarray}
&& \frac{d}{dt}
\rho_{0}=-\rho_{0}\Gamma_{\mathrm{abs}}+\rho_{\mathrm{X}}\Gamma_{\mathrm{rad}}
\\
&& \frac{d}{dt}
\rho_{\mathrm{X}^{\ast}}=\rho_{0}\Gamma_{\mathrm{abs}}-\rho_{\mathrm{X}^{\ast}}\gamma_{\mathrm{rel}}
\\
&&
\frac{d}{dt}\rho_{\mathrm{X}}=\rho_{\mathrm{X}^{\ast}}\gamma_{\mathrm{rel}}-\rho_{X}\Gamma_{\mathrm{rad}}
\label{eqn-rateeqn}
\end{eqnarray}

\noindent for steady-state, $d\rho_i/dt=0$. Here $\rho_i$ denotes
the population of state $| \mathrm{i}\rangle$ with $\rho_0 +
\rho_{\mathrm{X}^{\ast}}+ \rho_{\mathrm{X}} = 1 $, and
$\Gamma_{\mathrm{abs}}$, $\gamma_{\mathrm{rel}}$,
$\Gamma_{\mathrm{rad}}$ are the absorption, relaxation and
radiative recombination rates, respectively. For the population of
the exciton ground state $| \mathrm{X}\rangle$ we find:

\begin{eqnarray}
&&
\rho_{\mathrm{X}}=\frac{\Gamma_{\mathrm{abs}}}{\Gamma_{\mathrm{rad}}+(\Gamma_{\mathrm{abs}}/\gamma_{\mathrm{rel}})(\Gamma_{\mathrm{rad}}+\gamma_{\mathrm{rel}})}
\end{eqnarray}

\noindent which reduces in the limit $\gamma_{\mathrm{rel}} \gg
\Gamma_{\mathrm{rad}}$ to

\begin{eqnarray}
&&
\rho_{\mathrm{X}}=\frac{\Gamma_{\mathrm{abs}}}{\Gamma_{\mathrm{rad}}+\Gamma_{\mathrm{abs}}}.
\label{eqn-rho-X}
\end{eqnarray}

\noindent The PL intensity is given by $I_{\mathrm{PL}} = \eta \,
\Gamma_{\mathrm{rad}}\, \rho_{\mathrm{X}}$ with the experimental
detection quantum efficiency $\eta$. Normalizing the PL intensity
by the constant value in saturation
$I_{\mathrm{PL}}^{\mathrm{max}}$ we obtain using
Eq.~\ref{eqn-rho-X}:

\begin{eqnarray}
 \nonumber \frac{I_{\mathrm{PL}}}{I_{\mathrm{PL}}^{\mathrm{max}}}&=&\frac{1}{1+\beta}
\, ; \\ \beta&=&
\frac{\Gamma_{\mathrm{rad}}}{\Gamma_{\mathrm{abs}}} .
\label{eqn-I-PL-norm}
\end{eqnarray}

\noindent The saturation is uniquely determined by the ratio of
the radiative rate to the absorption rate. In the following we
relate $\beta$ to parameters which we access directly in our
experiments.

In the linear response, i. e. as long as the \textit{excited}
state transition is not saturated by the laser, the absorption
rate is proportional to the incident laser power
$P_{\mathrm{las}}$. The proportionality factor is given by the
ratio of the absorption cross-section $\Sigma_{\mathrm{X}^*}$ of
the $| 0\rangle \rightarrow | \mathrm{X}^*\rangle$ transition to
the focal spot area $\mathcal{A}$ (with $\mathcal{A}=1.13 \,
\mathrm{FWHM}^2$ for a focussed Gaussian beam; in our experiment
$\mathrm{FWHM}\simeq\lambda$). Expressing the laser power in terms
of the incident photon flux (i. e. the number of photons per
second)
$\dot{N}_{\mathrm{ph}}=P_{\mathrm{las}}/(\hbar\omega_{\mathrm{ph}})$
we obtain for the absorption rate:

\begin{equation}
\Gamma_{\mathrm{abs}}=\frac{\Sigma_{\mathrm{X}^*}}{\mathcal{A}}\dot{N}_{\mathrm{ph}}.
\label{eqn-Gamma-abs}
\end{equation}

\noindent We estimate the absorption cross-section of the excited
state using the radiatively limited scattering cross-section
$\Sigma_{\mathrm{X}}$ of the fundamental exciton transition $|
0\rangle \rightarrow | \mathrm{X}\rangle$ with transition
wavelength $\lambda/n_{r}$ ($n_{r}$ is the refractive index).
According to the optical theorem the scattering cross-section of a
radiatively limited transition is given by \cite{Jackson}:

\begin{equation}
\Sigma_{\mathrm{X}}=\frac{3}{2\pi}\left(\frac{\lambda}{n_{r}}\right)^2.
\label{eqn-scatter}
\end{equation}

\noindent Assuming that the oscillator strengths per atom are
equal for the ground and excited states, we have to take into
account that the excited state is spatially more extended. For a
one-dimensional harmonic potential the confinement length grows
with the quantum number $\nu$ as
$\sigma_{\mathrm{c}}^{\nu}=\sigma_{\mathrm{c}}^0\sqrt{2\nu+1}$. On
the other hand, the excited state is broadened by nonradiative
relaxation with rate $\gamma_{\mathrm{rel}}$. Thus we have for the
$\nu-$th excited state:

\begin{equation}
\Sigma_{\mathrm{X}^*}=\sqrt{2\nu+1}\left(\frac{\Gamma_{\mathrm{rad}}}{\gamma_{\mathrm{rel}}}\right)\Sigma_{\mathrm{X}}.
\label{eqn-scatter-XX}
\end{equation}

\noindent Combining Eq.~\ref{eqn-I-PL-norm}, \ref{eqn-Gamma-abs},
\ref{eqn-scatter}, and \ref{eqn-scatter-XX} we obtain:

\begin{equation}
\beta=\left(\frac{\mathcal{A}}{\Sigma_{\mathrm{X}}\sqrt{2\nu+1}}\right)\left(\frac{\gamma_{\mathrm{rel}}}{\dot{N}_{\mathrm{ph}}}\right).
\label{beta}
\end{equation}

\noindent All parameters quantifying $\beta$ in Eq.~\ref{beta} are
known from the experiment. Specifically, for the QD exciton in
Fig.s~3 and 4 of the manuscript, the relevant parameters are as
follows: the excitation resonance wavelength for the third excited
state $| \mathrm{X}^*\rangle$ with $\nu=3$ was
$\lambda_{\mathrm{exc}}=830$~nm, the emission of the fundamental
exciton state $| \mathrm{X}\rangle$ was detected in PL at
$\lambda_{\mathrm{PL}}=894$~nm, the radiative decay rate was
obtained from time-resolved PL as the inverse PL lifetime
$\Gamma_{\mathrm{rad}}=1/\tau_{\mathrm{PL}}$ with
$\tau_{\mathrm{PL}}=3.35$~ns, and the relaxation rate was deduced
from the linewidth of the PLE resonance as
$\gamma_{\mathrm{rel}}=6.1~\mathrm{meV}/\hbar$. With this complete
set of parameters we calculated the saturation behavior as a
function of laser power focussed to a spot area of
$\mathcal{A}=1.13\,\lambda_{\mathrm{exc}}^2$. The result of the
calculation is shown by the solid line in Fig.~4 of the
manuscript.

\end{document}